\documentclass[aps,prl,twocolumn,superscriptaddress]{revtex4-2}

\usepackage{amsmath,amssymb}
\usepackage{graphicx}
\usepackage{bm}
\usepackage{tikz}
\usetikzlibrary{arrows.meta,positioning,calc,decorations.pathmorphing}
\usepackage{placeins}

\begin{document}

\title{Wilson--Fisher Renormalization of Discrete Gravity--Capillary Wave Turbulence in Viscous Fluids}

\author{Jos\'e A. Santiago}
\affiliation{Departamento de Qu\'imica F\'isica, Universidad Complutense de Madrid, 28040 Madrid, Spain}
\affiliation{Departamento de Matem\'aticas Aplicadas y Sistemas Universidad Aut\'onoma Metropolitana Cuajimalpa
Vasco de Quiroga 4871, 05348 Cd. de M\'exico, Mexico.}

\author{Mikheil Kharbedia}
\affiliation{Departamento de Qu\'imica F\'isica, Universidad Complutense de Madrid, 28040 Madrid, Spain}
\affiliation{Translational Biophysics, Instituto de Investigaci\'on Sanitaria Hospital 12 de Octubre, 28041 Madrid, Spain}
\altaffiliation{Present address: Advanced Research Center for Nanolithography (ARCNL), Science Park 106, 1098 XG Amsterdam, The Netherlands}

\author{Basilio J. Garc\'ia}
\affiliation{Departamento de F\'isica Aplicada, Universidad Aut\'onoma de Madrid, 28049 Madrid, Spain}

\author{Francisco Monroy}
\email{monroy@ucm.es}
\affiliation{Departamento de Qu\'imica F\'isica, Universidad Complutense de Madrid, 28040 Madrid, Spain}
\affiliation{Translational Biophysics, Instituto de Investigaci\'on Sanitaria Hospital 12 de Octubre, 28041 Madrid, Spain}

\date{\today}

\begin{abstract}
We report an experimental realization of Wilson--Fisher renormalization in driven surface-wave turbulence across Newtonian fluids spanning nearly six decades in Reynolds number. Discrete capillary and gravity turbulence define two universality classes selected by interaction topology: triadic resonances for capillary waves and effectively tetradic scattering for gravity waves. Navier--Stokes viscosity is the relevant perturbation that renormalizes spectral transfer and terminates the cascade. The resulting framework predicts the Kolmogorov cutoff from the balance of nonlinear transfer and viscous damping, and Reynolds scaling of the integrated inertial spectral weight. Laser Doppler vibrometry quantitatively confirms these renormalized scaling laws, establishing discrete gravity--capillary turbulence as a tunable laboratory for nonequilibrium crossover criticality.
\end{abstract}

\maketitle

Universality in turbulence reflects the separation between inertial transfer and dissipation \cite{LandauLifshitzFM}. In fully developed hydrodynamic turbulence, Kolmogorov's phenomenology identifies an inertial cascade terminated at small scales by viscous dissipation \cite{Kolmogorov1941a,Frisch1995}. This scale separation can be phrased in Wilsonian terms \cite{WilsonKogut1974}: coarse graining integrates out fast degrees of freedom to yield scale-dependent effective dynamics, while relevant perturbations—viscosity as the paradigmatic example, control departures from scale invariance \cite{Fisher1974}. A long-standing renormalization-group (RG) program for Navier–Stokes (NS) turbulence follows this logic \cite{Wyld1961,ForsterNelsonStephen1977,EyinkGoldenfeld1994}, using RG flows to predict renormalized transport \cite{YakhotOrszag1986} and functional scaling \cite{SmithWoodruff1998,Canet2022FRGTurbulence}. By contrast, wave turbulence (WT) theory is largely formulated around inviscid fixed points \cite{Zakharov1992,Nazarenko2011} and, despite renormalization procedures for interactions and spectra \cite{EyinkGoldenfeld1994,Canet2022FRGTurbulence}, lacks a Wilsonian organization of viscous crossover flows in experimentally accessible RG coordinates. Here we adopt a Wilson–Fisher (WF) viewpoint, where relevant perturbations control the approach to scale-invariant fixed points and generate observable crossover scaling \cite{WilsonFisher1972}.

\begin{figure}[htb!]
  \centering
  \includegraphics[width=\columnwidth]{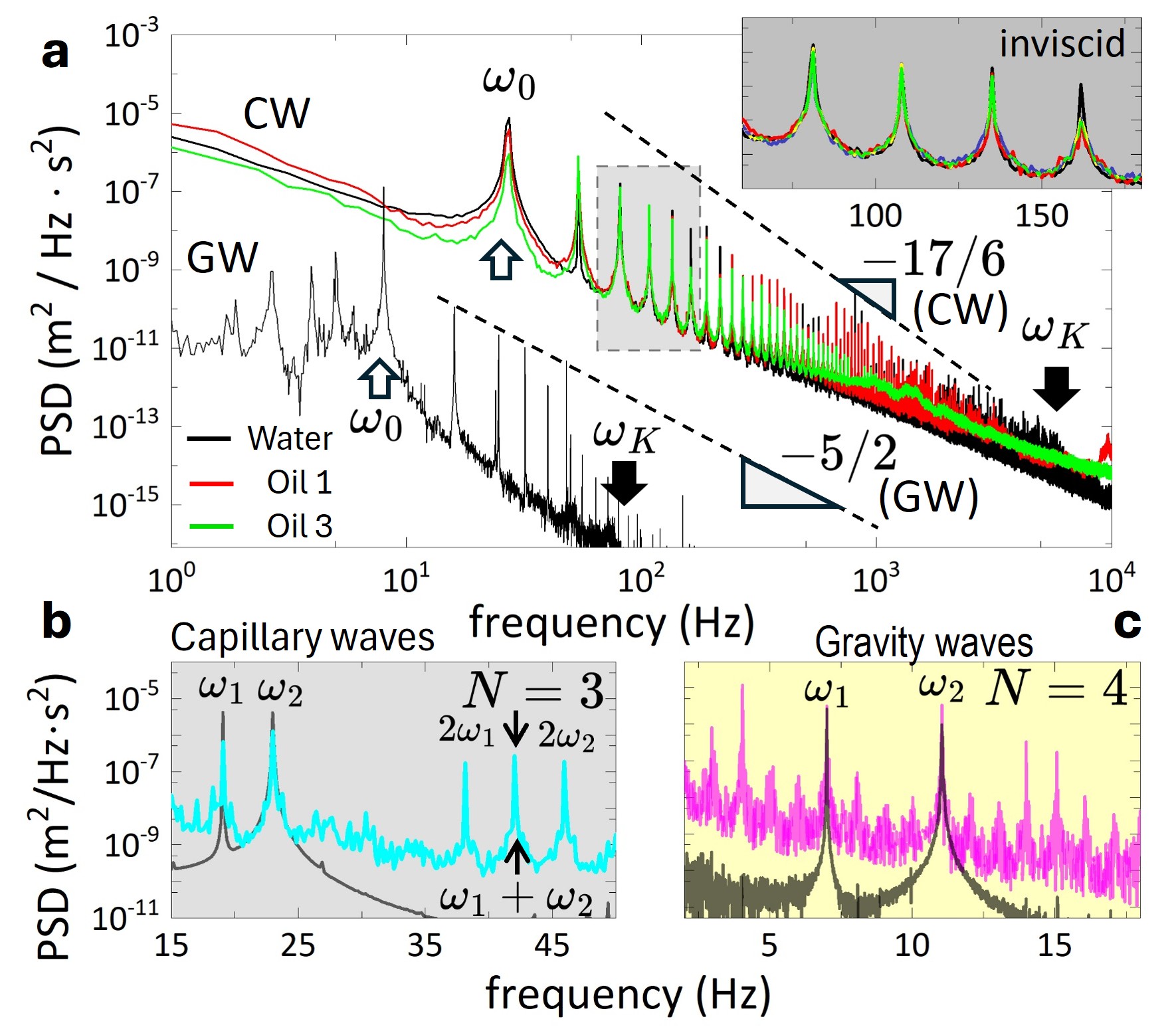}
\caption{
Frequency-domain signatures and inertial scaling of discrete surface-wave turbulence (DSWT) in viscous fluids.
(a) Broadband surface-velocity power spectral density $S(\omega)$ showing the capillary-wave (CW; top) and gravity-wave (GW; bottom) cascades for water and high-viscosity oils (Oil~1: $10$, Oil~3: $10^{3}$ times the viscosity of water). Open arrows mark the excitation frequency $\omega_0$, filled ones the Kolmogorov cutoff $\omega_K$. Dashed guide lines indicate KZ scaling: $S(\omega)\sim\omega^{-17/6}$ (CW) and $S(\omega)\sim\omega^{-5/2}$ (GW). Inset: discrete inviscid CW modes persist across the inertial range, essentially independent of bulk viscosity.
(b) CW spectrum illustrating three-wave ($N=3$) resonant mixing, with peaks at $\omega_1$, $\omega_2$, $2\omega_1$, $2\omega_2$, and $\omega_1+\omega_2$.
(c) GW spectrum illustrating four-wave ($N=4$) transfer, with dominant peaks at $\omega_1$, $\omega_2$ and dismutation resonances.
}

\label{fig:capillary_gravity_scaling}
\end{figure}

Nonlinear surface waves, including gravity (GW) and capillary (CW) modes, provide a canonical setting where dispersion selects the resonant interaction channels~\cite{Dias&Kharif1999}.
In the weak regime, they support Kolmogorov--Zakharov (KZ) fixed points with surface-velocity spectra $S(\omega)\sim\omega^{-p}$ \cite{Zakharov1992,Nazarenko2011}.
As summarized in Fig.~\ref{fig:capillary_gravity_scaling}, CW and GW turbulence define two universality classes set by interaction topology: CW is governed by exact three-wave resonances, whereas deep-water GW forbids triads and is controlled by effective four-wave (tetradic) scattering via quasi-resonant quartets~\cite{Dias&Kharif1999}.
In both classes, spectral discreteness persists across the inertial range, while viscosity terminates the cascade at the Kolmogorov cutoff $\omega_K$—dispersion selects the class, dissipation sets exit.

In this Letter, we develop and test a Wilsonian framework for discrete surface-wave turbulence (DSWT) under monochromatic forcing at $\omega_0$, where dissipation controls the crossover dynamics of the GW and CW cascade response (Fig.~\ref{fig:capillary_gravity_scaling}). We cast this behavior in a Wilson--Fisher (WF) picture \cite{WilsonFisher1972,WilsonKogut1974}: interaction topology selects conservative fixed points, while Navier--Stokes viscosity is the relevant perturbation that renormalizes transfer and terminates the inertial interval at a viscous (Kolmogorov) cutoff $\omega_K$ (Fig.~\ref{fig:capillary_gravity_scaling}a). Using laser Doppler vibrometry (LDV) \cite{Cobelli2009PRL,Kharbedia2025}, we first observe a robust capillary-wave inertial cascade with the expected KZ decay of the surface-velocity PSD, $S(\omega)\sim\omega^{-17/6}$ for $\omega>\omega_0$ \cite{Nazarenko2011}, persisting without measurable resonance broadening over nearly three decades of viscosity (Fig.~\ref{fig:capillary_gravity_scaling}a; inset). This invariance is consistent with exact three-wave resonances in dispersive CWs, $\mathbf{k}_1+\mathbf{k}_2=\mathbf{k}_3$ and $\omega_1+\omega_2=\omega_3$ (i.e., detuning $\Delta\equiv\omega_1+\omega_2-\omega_3=0$), indicating strictly conservative triadic coupling (Fig.~\ref{fig:capillary_gravity_scaling}b), as described in Kartashova’s theory of discrete wave turbulence \cite{Kartashova2007PRL,Kartashova2009EPL}. By contrast, in the GW branch we observe a direct inertial decay consistent with the weak-turbulence prediction $S(\omega)\sim\omega^{-5/2}$ for $\omega>\omega_0$ \cite{Dyachenko2004PRLGravity}. An enhanced subharmonic response appears below $\omega_0$ (Fig.~\ref{fig:capillary_gravity_scaling}a, bottom), consistent with weak reverse transfer and \emph{dismutation} signatures \cite{Dyachenko2004PRLGravity,Kartashova2009EPL}. In dismutation, two pumped modes $(\omega_1,\omega_2)$ scatter into a frequency-split quartet, generating one daughter above the pumped band and one below (Fig.~\ref{fig:capillary_gravity_scaling}c). Intraband products may also arise via harmonics/subharmonics of pumping. This is consistent with quasi-elastic tetradic scattering, $\mathbf{k}_1+\mathbf{k}_2=\mathbf{k}_3+\mathbf{k}_4$, where the frequency condition is quasi-resonant, $\omega_1+\omega_2=\omega_3+\omega_4+\Delta$ with small detuning $\Delta\neq0$, reflecting resonance broadening for underdispersive gravity waves $\omega_G\propto k^{1/2}$ \cite{Dyachenko2004PRLGravity,Nazarenko2011} and inexact resonance clustering \cite{Kartashova2007PRL,Kartashova2009EPL}.

Our analysis exploits a systematic dataset acquired across Newtonian fluids with variable surface tension $\sigma$ and density $\rho$, spanning more than three decades in dynamic viscosity $\eta$, including water and silicone oils (Table~I). The surface-velocity spectrum is driven monochromatically at angular frequency $\omega_0$ with displacement amplitude $A$, measured noninvasively by laser Doppler vibrometry (LDV) \cite{Cobelli2009PRL,Kharbedia2025}. Across fluids, the kinematic viscosity $\nu=\eta/\rho$ spans $\sim10^{4}$ and, over the explored $\omega_0$ range, the LDV-resolved forcing velocity $U_0\equiv A\omega_0$ spans $\sim10^{3}$, enabling a wide sweep of Reynolds number, ${\rm Re}\equiv U_0\Lambda_0/\nu$, while keeping the injection scale $\Lambda_0$ comparable. The drive selects the forcing wavenumber $k_0$ via the linear dispersion relation (GW/CW); we define the injection length $\Lambda_0\equiv 2\pi/k_0$, and the injected kinetic-energy density $E=\tfrac12\rho U_0^2$. In all fluids, both CW and GW branches display direct KZ-type cascades for $\omega>\omega_0$, extending over a discrete inertial interval that terminates at the viscous cutoff $\omega_K$ (Fig.~\ref{fig:capillary_gravity_scaling}a). These experimentally resolved features motivate a WF organization in terms of two reduced RG coordinates: the ultraviolet-to-injection viscous-rate ratio $\bar{\Omega}\equiv\omega_K/(\nu k_0^2)$ and the integrated inertial spectral weight $\Sigma_{\mathrm{PSD}}=\int_{\omega_0}^{\omega_K} S(\omega),d\omega$, which quantifies the global cascade response to forcing. In addition, GW spectra exhibit subharmonic dismutation signatures (Fig.~\ref{fig:capillary_gravity_scaling}c) and enhanced response below $\omega_0$ (Fig.~\ref{fig:capillary_gravity_scaling}a, bottom), consistent with weak reverse transfer and a much smaller spectral weight than CW, $\Sigma_{\mathrm{PSD}}^{(G)}\approx10^{-3}\Sigma_{\mathrm{PSD}}^{(C)}$, yielding two energetically well separated, non-overlapping responses. This WF structure predicts the scaling of $\bar{\Omega}$ with $E$ and the Reynolds renormalization of the reduced cascade response $\bar\Sigma\equiv\Sigma_{\mathrm{PSD}}/(\Lambda_0\omega_0)^2$, yielding distinct RG exponents for triadic capillary and tetradic gravity cascades. Unlike WT treatments focused on fixed-point spectra, promoting $\bar{\Omega}$ and $\bar\Sigma$ to RG coordinates enables direct observation of crossover trajectories and organizes DSWT into two topology-defined attraction basins.

\begin{table}[t]
\caption{\label{tab:fluids}
Material parameters of the fluids used in the DSWT experiments, ordered by increasing kinematic viscosity $\nu$.
Surface tension $\sigma$, density $\rho$, and dynamic viscosity $\eta$ are shown to emphasize that kinematic viscosity acts as the primary relevant perturbation in the RG flow.
}
\centering
\begin{ruledtabular}
\begin{tabular}{lcccc}
Fluid &
$\sigma$ (mN\,m$^{-1}$) &
$\eta$ (mPa\,s) &
$\rho$ (g\,cm$^{-3}$) &
$\nu$ (mm$^2$\,s$^{-1}$)
\\ \hline
Mercury            & 487 & 1.6   & 13.6 & 0.12 \\
n-Hexane           & 18  & 0.30  & 0.70 & 0.43 \\
Chloroform         & 27  & 0.54  & 1.50 & 0.47 \\
CCl$_4$            & 27  & 1.0   & 1.60 & 0.63 \\
Water (Milli-Q)    & 72  & 0.82  & 1.00 & 0.82 \\
Pb(NO$_3$)$_2$ (sat.) & 90  & 2.5   & 1.30 & 2.3 \\
Oil 1 (RTM8)       & 32  & 10    & 0.80 & 13   \\
Oil 2 (RTM14)      & 32  & 100   & 0.80 & 125  \\
Glycerol           & 63  & 1410  & 1.30 & 1100 \\
Oil 3 (RTM18)      & 32  & 1000  & 0.80 & 1250 \\
\end{tabular}
\end{ruledtabular}
\end{table}

\paragraph{Wilson--Fisher renormalization framework.—}

In the inviscid limit, surface-wave turbulence is governed by resonant interactions constrained by dispersion, leading to scale-invariant KZ--spectra in the inertial interval \cite{Zakharov1992,Nazarenko2011}. In Wilson--Fisher (WF) language, these KZ states act as conservative fixed points, while the interaction topology---triadic for capillary waves and effectively tetradic for gravity waves (Fig.~1), selects the universality class \cite{Zakharov1967Capillary,Dyachenko2004PRLGravity}. To render the RG structure explicit, we introduce a running frequency scale $\omega$ and define as dimensionless coupling the squared wave steepness
\begin{equation}
g(\omega)\equiv \epsilon(\omega)^2 \sim k(\omega)^2\,S(\omega)\,\omega ,
\label{eq:running_coupling}
\end{equation}
where $S(\omega)$ is the surface-velocity spectrum and $k(\omega)$ follows from the linear dispersion relation \cite{Dias&Kharif1999}. Writing $S(\omega)\sim \omega^{-p}$, one has $g(\omega)\sim \omega^{y_g}$ with
\begin{equation}
y_g \equiv 2\alpha-(p-1),\qquad k(\omega)\sim \omega^{\alpha},
\label{eq:yg_def}
\end{equation}
and the RG ``time'' $\ell\equiv \ln(\omega/\omega_0)$. The coarse-graining flow is then captured at scaling level by a WF beta function \cite{WilsonKogut1974,Wegner1972} (see SM for details),
\begin{equation}
\beta(g)\equiv \frac{dg}{d\ell}=y_g\,g - B_N\,g^{\,N-1},
\label{eq:beta_wf}
\end{equation}
where $B_N>0$ encodes nonlinear saturation and the interaction topology enters through the order $N$ of the resonant process. \textit{(End Matter connects the topology of the interaction vertex to the saturation term.)}
Besides the Gaussian fixed point $g^\star=0$, Eq.~(\ref{eq:beta_wf}) admits a nontrivial stable fixed point $g^\star\sim (y_g/B_N)^{1/(N-2)}$ corresponding to the KZ inertial state. In the weak regime we use $p=17/6$ (capillary) and $p=5/2$ (gravity) \cite{Zakharov1992,Nazarenko2011}.

\textit{Kolmogorov scale.} NS--viscosity explicitly breaks the inviscid scale invariance and provides the RG-relevant perturbation that drives the system away from each conservative fixed point \cite{Wyld1961}. For small-amplitude surface waves, linear viscous hydrodynamics yields a damping rate $\gamma_\nu(k)\propto \nu k^2$ and a viscous time $\tau_\nu(k)\approx(\nu k^2)^{-1}$ \cite{Lamb1932,LandauLifshitzFM}. This Kolmogorov UV-relevant dissipative operator is common to both CW/GW turbulence because it is inherited directly from the bulk NS--term $\nu\nabla^2\bm{u}$ \cite{LandauLifshitzFM}; the only regime dependence enters through the nonlinear transfer time, which is set by the interaction topology.

\textit{Fixed points: universality classes.} The inertial interval terminates at the experimentally resolved Kolmogorov cutoff $\omega_K$ (Fig.~1a), selected by the crossover condition $\tau_{\mathrm{nl}}(\omega_K)\approx\tau_\nu(\omega_K)$, in direct analogy with Kolmogorov's dissipation scale \cite{Kolmogorov1941a,Frisch1995,Nazarenko2011}. To make explicit how interaction topology enters this cutoff, we relate the nonlinear rate to the inertial spectrum at scaling level. For a narrow band around $\omega$, the typical wave amplitude satisfies $a_\omega^2\sim S(\omega)\,\omega$, while the dimensionless nonlinearity is the steepness $\epsilon(\omega)\sim k(\omega)\,a_\omega$. In weak turbulence, the kinetic transfer rate generated by an $N$-wave resonant process scales as $\tau_{\mathrm{nl}}^{-1}(\omega)\sim \omega\,\epsilon(\omega)^{2(N-2)}$ \cite{Zakharov1992}, so that
\begin{equation}
\tau_{\mathrm{nl}}^{-1}(\omega)\sim \omega\,[k(\omega)^2\,S(\omega)\,\omega]^{\,N-2},
\quad
N=
\begin{cases}
3 & \text{(CW triads)}\\
4 & \text{(GW tetrads)}
\end{cases}
\label{eq:bridge_rate}
\end{equation}
(details and prefactors are given in the SM). Because $\tau_\nu^{-1}\approx\nu k^2$ is inherited universally from Navier--Stokes, while $\tau_{\mathrm{nl}}$ depends on the interaction order $N$ and the KZ exponent $p$, the ultraviolet cutoff scaling is class-dependent even though the viscous substrate is universal.

\begin{figure}[t]
    \centering
    \includegraphics[width=\columnwidth]{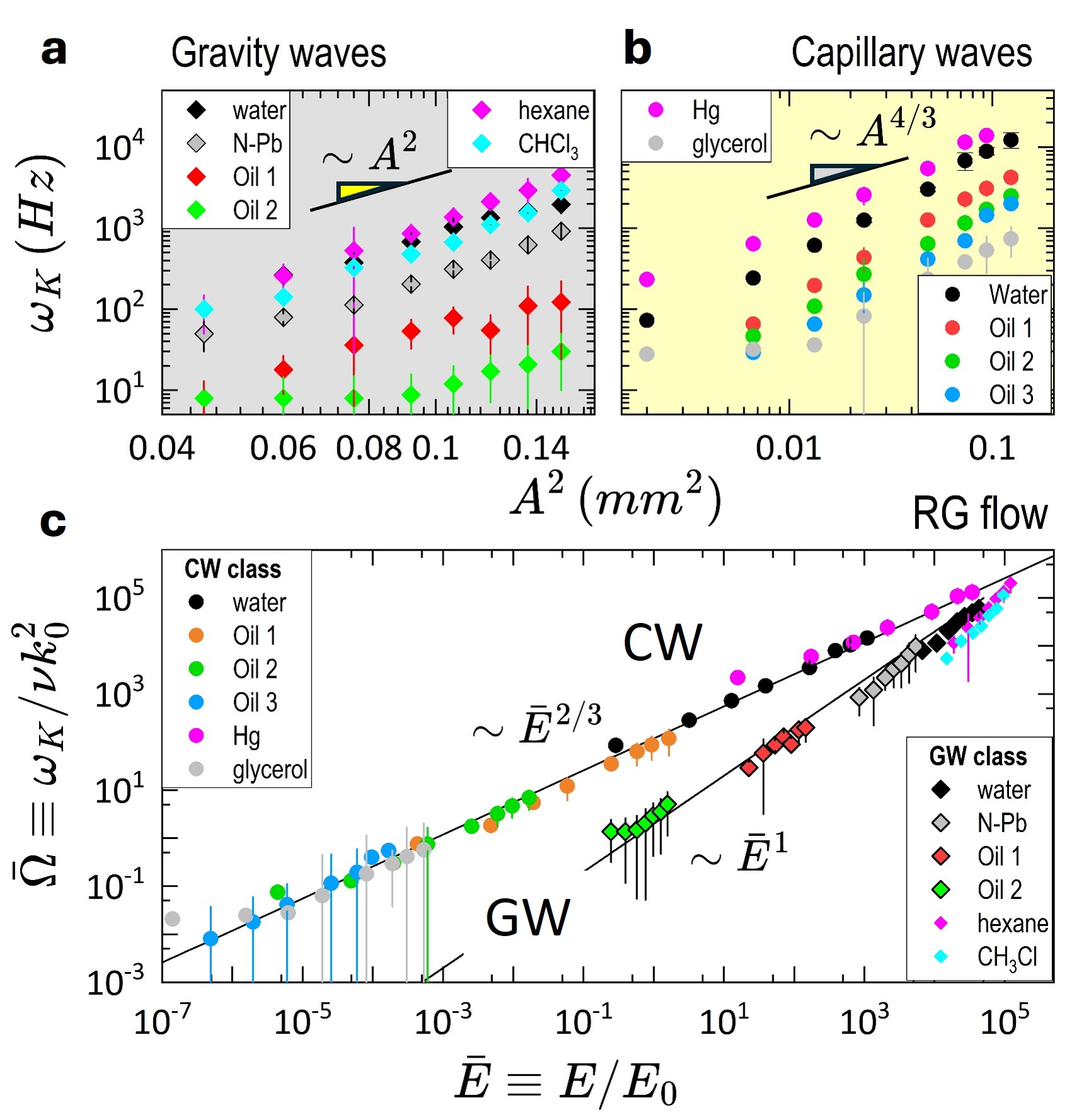}
    \caption{
    Kolmogorov cutoff as a Wilson--Fisher ultraviolet scale.
    (a) GW ($\omega_0=8\,\mathrm{Hz}$): $\omega_K$ vs forcing amplitude $A$.
    (b) CW ($\omega_0=27\,\mathrm{Hz}$): $\omega_K$ vs $A$.
    In both regimes, $\omega_K$ grows monotonically with forcing, defining the Kolmogorov exit scale.
    (c) Wegner scaling of the ultraviolet coordinate $\bar{\Omega}_\nu\equiv \omega_K/(\nu k_0^2)$ with $k_0=2\pi/\Lambda_0$ and
    $\Lambda_0$ the dispersion-selected forcing length, versus reduced injected energy $\bar{E}\equiv E/E_0$, with
    $E=\rho (A\omega_0)^2/2$ and $E_0=\rho(\nu/\Lambda_0)^2$ the viscous kinetic-energy scale at $\Lambda_0$ (see SM).
    Data collapse onto topology-selected branches: $\bar{\Omega}_\nu\sim \bar{E}^{1}$ (GW, $N=4$) and $\bar{\Omega}_\nu\sim \bar{E}^{2/3}$ (CW, $N=3$).
    Solid lines: Wilson--Fisher predictions from Eq.~(\ref{eq:omegamax}).
    }
    \label{fig:RG_cutoff}
\end{figure}

We take the injected kinetic-energy density at the driven mode as $E=\tfrac12\rho U_0^2$ with $U_0\equiv A\omega_0$, and interpret the measured $\omega_K$ as the ultraviolet RG scale at which the flow exits the inertial basin. Furthermore, we define a Reynolds-like reduced injected energy
$\bar E \equiv E\,\Lambda_0^{2}/(\rho \nu^{2})
= \tfrac12 (U_0\Lambda_0/\nu)^2=\tfrac12\,{\rm Re}^2$,
which provides a dimensionless forcing coordinate at the injection scale.
The WF crossover framework then yields the experimentally testable cutoff laws for each universality class
\begin{equation}
\bar \Omega_\nu^{(CW)} \sim \bar{E}^{2/3}\ (N=3),\qquad \bar \Omega_\nu^{(GW)} \sim \bar{E}^{1}\ (N=4).
\label{eq:omegamax}
\end{equation}

\paragraph{Cascade energy: Reynolds renormalization.—}
Beyond the ultraviolet coordinate $\bar{\Omega}_\nu\equiv \omega_K/(\nu k_0^2)$, a second RG observable is the integrated inertial spectral weight
$\Sigma_{\mathrm{PSD}}=\int_{\omega_0}^{\omega_K}S(\omega),d\omega$, which measures the total velocity variance accumulated in the direct cascade between the pump and the UV-exit scale $\omega_K$.
The corresponding raw response $\Sigma_{\mathrm{PSD}}(U_0)$ is shown in Supplementary Fig.~S1: it increases monotonically with forcing velocity $U_0=A\omega_0$ in both universality classes, yet remains separated by $\sim$3 decades, reflecting a much weaker GW cascade.
To expose universal renormalization under injection, we define the reduced response $\bar\Sigma_\omega({\rm Re})\equiv\Sigma_{\mathrm{PSD}}/(\Lambda_0\omega_0)^2$ with $\Lambda_0=2\pi/k_0$ and ${\rm Re}\equiv U_0\Lambda_0/\nu$.
Within the WF crossover framework, $\bar\Sigma$ obeys a Wegner scaling form $\bar\Sigma({\rm Re})\sim{\rm Re}^{\kappa_N}\mathcal{F}_N({\rm Re}/{\rm Re}_\times)$ \cite{Wegner1972}; in the asymptotic regime, the renormalization exponent is fixed by interaction topology and by the fact that $\bar\Sigma_\omega$ probes a quadratic (second-moment) cascade response (see SM), giving $\kappa_N=2/(N-1)$ and thus
\begin{equation}
\bar\Sigma_\omega^{(CW)} \sim {\rm Re}^{1}\ (N=3),\quad \bar\Sigma_\omega^{(GW)} \sim {\rm Re}^{2/3}\ (N=4),
\label{eq:Sigma_Re}
\end{equation}
quantitatively confirmed by the collapse in Fig.~3.

\begin{figure}[t]
    \centering
    \includegraphics[width=\columnwidth]{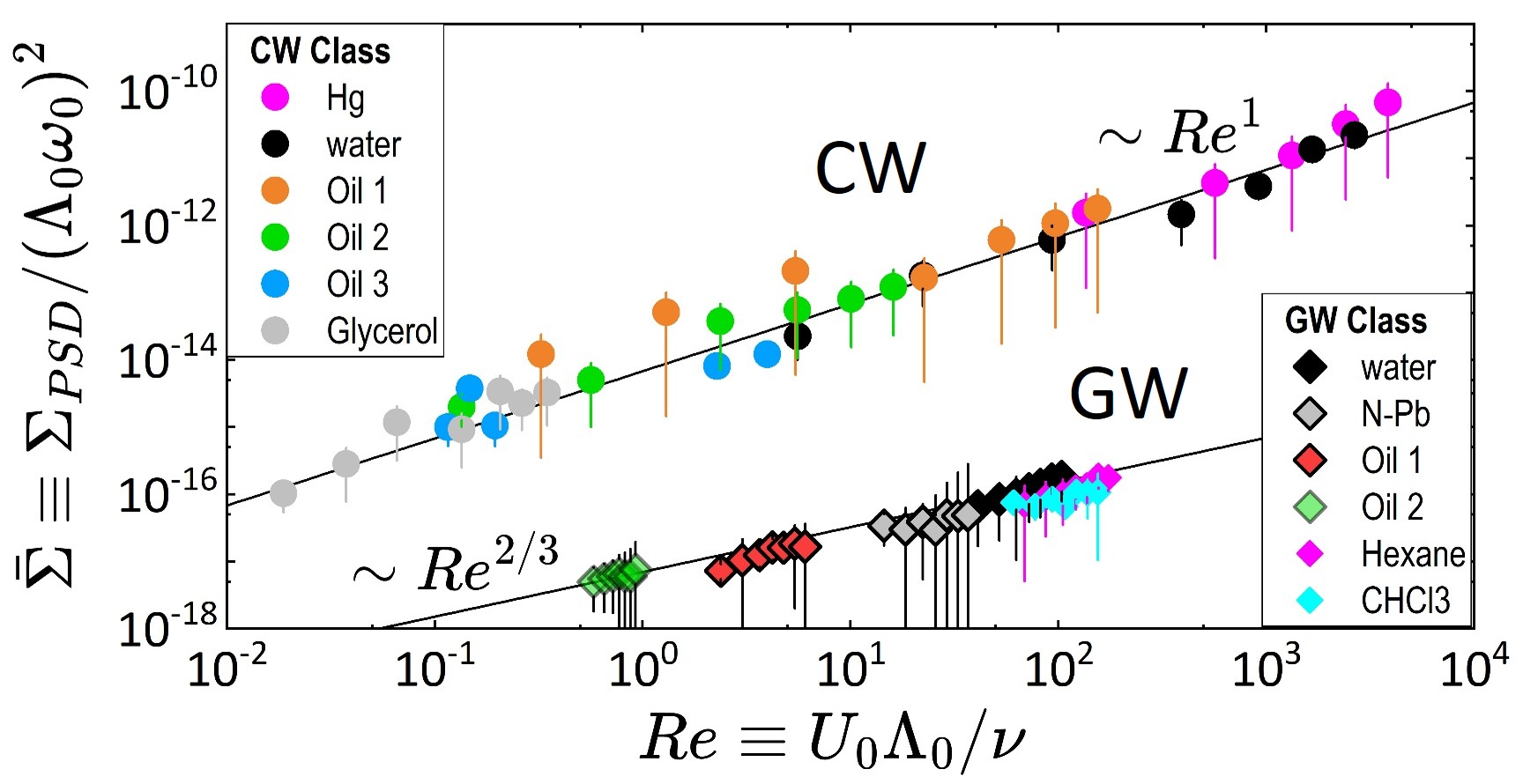}
    \caption{
Reynolds-number renormalization of the cascade response.
Reduced inertial spectral weight $\bar\Sigma\equiv\Sigma_{\mathrm{PSD}}/(\Lambda_0\omega_0)^2$ versus ${\rm Re}\equiv U_0\Lambda_0/\nu$ across all fluids, revealing two energetically separated universality classes: capillary-wave (CW, upper) and gravity-wave (GW, lower) turbulence.
Data follow WF/Wegner scaling and collapse onto topology-controlled power laws
$\bar\Sigma_\omega\sim{\rm Re}^{2/(N-1)}$:
$\bar\Sigma\sim{\rm Re}^{1}$ for CW ($N=3$) and $\bar\Sigma_\omega\sim{\rm Re}^{2/3}$ for GW ($N=4$)
(solid lines: WF predictions from Eq.~(\ref{eq:Sigma_Re})), demonstrating interaction-topology renormalization of the global cascade response (SM; End Matter).
}
    \label{fig:RG_Re_scaling}
\end{figure}

\paragraph{Discussion.—}
The central outcome is a genuinely Wilsonian organization of discrete surface-wave turbulence (DSWT): the experimentally accessible coordinates $(\bar{\Omega}_\nu,\bar\Sigma_\omega)$ provide a two-parameter RG chart in which monochromatically forced DSWT separates into two attraction basins, each governed by a distinct conservative KZ fixed point selected by interaction topology ($N=3$ for CW and $N=4$ for GW).
\emph{Conceptual punchline:} DSWT admits a Wilsonian RG constructed from directly measured exit and response coordinates, which organize the dynamics into topology-defined basins of attraction.
NS--viscosity plays the role of a single relevant perturbation that generates crossover trajectories and terminates the inertial flow at the exit scale $\omega_K$.
The simultaneous verification of the Wegner-like law $\bar{\Omega}_\nu(\bar E)$ (Fig.~2) and the Reynolds renormalization $\bar\Sigma_\omega({\rm Re})$ (Fig.~3) thus constitutes a direct experimental realization of WF--crossover criticality in a nonequilibrium cascade \textit{(End Matter, Fig.~4 gives a diagrammatic representation of the triadic and tetradic interaction channels)}.
Beyond providing a controlled laboratory for RG ideas long pursued in NS turbulence, DSWT offers a tunable setting where the universality class is switched by dispersion, while the relevant perturbation is continuously dialed by viscosity.
Finally, the two RG coordinates probe complementary pieces of the WF structure: the ultraviolet exit coordinate $\bar{\Omega}_\nu(N,y_g)$ is fixed by the relevance of the steepness coupling under $\omega$--coarse graining, set by topology $N$ and the engineering exponent $y_g$ in $\beta(g)=y_g g - B_N g^{,N-1}$, whereas the integrated quadratic response $\bar\Sigma_\omega(N,B_N;g_0^2)$ is set by nonlinear saturation, hence by the same topology through $B_N$ and the combinatorial power $N-1$.

\paragraph{Conclusion.—}
Discrete gravity and capillary wave turbulence realize two Wilson--Fisher universality classes with topology-selected fixed points and viscosity-controlled crossover.
The resulting RG coordinates $(\bar{\Omega},\bar\Sigma)$ are directly measurable and obey distinct renormalized scaling laws, establishing DSWT as a tunable platform for nonequilibrium Wilsonian criticality.\\

\begin{acknowledgments}
We thank Prof. M.~G.~Velarde for insightful discussions on the role of dimensionless numbers in turbulence. This work was supported by the Agencia Estatal de Investigación (AEI, Spain) under Grants TED2021-132296B-C52 and CPP2024-011880, and by Fundación BBVA--Fundamentos 2025 (to F.M.), and by the Secretaría de Ciencia, Humanidades, Tecnología e Innovación (SECIHTI, Mexico) under Grant CBF-2025-I-4418 (to J.A.S.).
\end{acknowledgments}

\bibliographystyle{apsrev4-2}
\bibliography{dswt_rg}

\clearpage


\section*{End Matter: Diagrammatic origin of topology-dependent RG flows}

\begin{figure}[!h]
\centering
\begin{tikzpicture}[
    font=\small,
    line/.style={line width=0.65pt},
    vtx/.style={circle, fill=black, inner sep=1.4pt},
    lab/.style={inner sep=12pt, align=left},
    >=Latex
]

\def\x{0}

\node[lab] at (\x, 5.40) {\textbf{(a) Capillary waves (CW): triads ($N=3$)}};

\coordinate (Va) at (\x, 3.95);
\node[vtx] at (Va) {};
\draw[line] (Va) -- ++(-2.10, 0.85) node[lab, left] {$1$};
\draw[line] (Va) -- ++(-2.10,-0.85) node[lab, left] {$2$};
\draw[line] (Va) -- ++( 2.25, 0.00) node[lab, right] {$3$};

\node[lab] at (\x+2.55, 4.55)
{$\mathbf{k}_1+\mathbf{k}_2=\mathbf{k}_3$\\[-1pt]
$\omega_1+\omega_2=\omega_3$};

\node[lab] at (\x-2.35, 2.55) {\textit{self-renormalization}};

\coordinate (Ain)  at (\x-2.05, 2.35);
\coordinate (Aout) at (\x+2.05, 2.35);
\draw[line] (Ain) -- (\x-1.05, 2.35);
\draw[line] (\x+1.05, 2.35) -- (Aout);

\coordinate (V1a) at (\x-0.75, 2.35);
\coordinate (V2a) at (\x+0.75, 2.35);
\node[vtx] at (V1a) {};
\node[vtx] at (V2a) {};
\draw[line] (V1a) -- (V2a);
\draw[line] (V1a) to[out=110,in=70] (\x, 2.85) to[out=250,in=290] (V2a);

\node[lab] at (\x+2.55, 2.35)
{$\delta g\sim g^{2}$\\[3pt]
$\beta(g)=y_g g-B_{3}g^{2}$};


\node[lab] at (\x, 0.95) {\textbf{(b) Gravity waves (GW): tetrads ($N=4$)}};

\coordinate (Vb) at (\x, -0.55);
\node[vtx] at (Vb) {};
\draw[line] (Vb) -- ++(-1.90, 0.95) node[lab, left]  {$1$};
\draw[line] (Vb) -- ++(-1.90,-0.95) node[lab, left]  {$2$};
\draw[line] (Vb) -- ++( 1.90, 0.95) node[lab, right] {$3$};
\draw[line] (Vb) -- ++( 1.90,-0.95) node[lab, right] {$4$};

\node[lab] at (\x-2.70, -0.25)
{$\mathbf{k}_1+\mathbf{k}_2=\mathbf{k}_3+\mathbf{k}_4$\\[-1pt]
$\omega_1+\omega_2=\omega_3+\omega_4+\Delta$};

\draw[line, decorate, decoration={snake, amplitude=0.60mm, segment length=2.0mm}]
(\x, -0.10) -- (\x, 0.55);
\node[lab] at (\x+0.60, 0.35) {$\Delta\neq 0$};

\node[lab] at (\x-2.35, -2.10) {\textit{self-renormalization}};

\coordinate (Bin)  at (\x-2.05, -2.30);
\coordinate (Bout) at (\x+2.05, -2.30);
\draw[line] (Bin) -- (\x-1.05, -2.30);
\draw[line] (\x+1.05, -2.30) -- (Bout);

\coordinate (V1b) at (\x-0.75, -2.30);
\coordinate (V2b) at (\x,      -2.30);
\coordinate (V3b) at (\x+0.75, -2.30);
\node[vtx] at (V1b) {};
\node[vtx] at (V2b) {};
\node[vtx] at (V3b) {};
\draw[line] (V1b) -- (V2b);
\draw[line] (V2b) -- (V3b);
\draw[line] (V1b) to[out=110,in=70] (\x, -1.80) to[out=250,in=290] (V3b);

\node[lab] at (\x+2.55, -2.30)
{$\delta g\sim g^{3}$\\[+3pt]
$\beta(g)=y_g g-B_{4}g^{3}$};

\end{tikzpicture}

\vspace{-2mm}
\caption{\textbf{Topology-dependent interaction vertices and WF saturation at fixed points.}
(a) Capillary-wave turbulence admits exact triads ($N=3$), yielding $\delta g\sim g^2$ and
$\beta(g)=y_g g-B_3 g^2$.
(b) Deep-water gravity-wave turbulence proceeds via effective tetrads ($N=4$) with finite resonance
broadening (detuning insertion $\Delta\neq 0$), yielding $\delta g\sim g^3$ and
$\beta(g)=y_g g-B_4 g^3$.
Interaction topology fixes the nonlinear term in $\beta(g)$ and defines distinct RG universality classes.}
\label{fig:EndMatter_Fig4}
\end{figure}
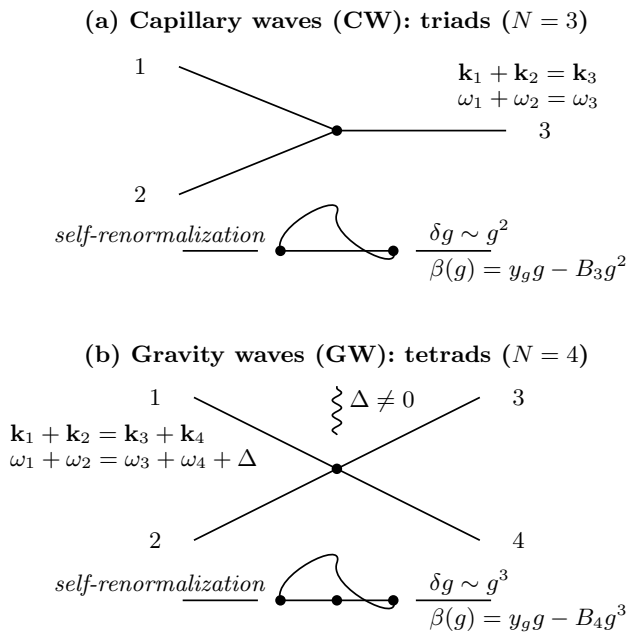

Weak wave turbulence can be described as resonant coupling between wave modes, encoded by effective interaction
vertices \cite{Zakharov1967Capillary,Nazarenko2011,Kartashova2007PRL}. Figure~\ref{fig:EndMatter_Fig4} summarizes
the minimal diagrammatic content needed here: the lowest-order resonant process fixes the interaction topology and
thus the universality class. Capillary waves (CW) admit exact three-wave resonances (triads, $N=3$). Deep-water
gravity waves (GW) forbid exact triads, so transfer proceeds through effective four-wave interactions (tetrads,
$N=4$), with a finite resonance width (schematically indicated by a detuning insertion $\Delta\neq 0$). Under
Wilsonian coarse graining, this topology controls the leading nonlinear saturation of the running coupling
$g(\omega)=\epsilon(\omega)^2$ and yields a Wilson--Fisher beta function $\beta(g)=y_g g - B_N g^{\,N-1}$, where the
power $N-1$ is fixed by the dominant vertex.


Here $y_g$ sets the linear tendency of $g$ under $\omega$-coarse graining, while $B_N$ sets the leading nonlinear
saturation strength. Figure~\ref{fig:EndMatter_Fig4} then encodes the key distinction: CW self-renormalizes at order
$g^2$ through the triadic channel, whereas GW self-renormalizes at order $g^3$ through the first available effective
tetradic channel, enabled by a finite resonance width (detuning insertion $\Delta\neq 0$). Experimentally, the RG
trajectory is traversed by tuning ${\rm Re}$, which controls the strength of viscous dissipation relative to forcing
at fixed driving frequency and injection geometry. Increasing ${\rm Re}$ weakens the viscous perturbation and
allows $g(\omega)$ to grow smoothly from the near-Gaussian regime toward the topology-selected KZ fixed point. This topology-controlled saturation governs the renormalization of the global cascade response in discrete surface wave turbulence (DSWT), yielding the distinct power laws $\bar\Sigma_\omega({\rm Re})$ in Fig.~3. Along the same flow, the inertial interval extends further into the ultraviolet before being arrested by viscosity, so the exit scale $\omega_K$, and hence renormalized $\bar\Omega_\nu(Re^{1/2})$, increases with forcing as captured by the WF/Wegner scaling in Fig.~2. The attraction basin is therefore selected kinematically by dispersion (which fixes the resonant topology), while dissipation, tuned by ${\rm Re}$, sets the crossover by controlling how far the flow proceeds along the corresponding RG trajectory toward its fixed point.

\paragraph{Relation to kinetic and discrete wave-turbulence theories.—} Classical wave-turbulence (WT) theory is formulated in terms of kinetic equations that describe stationary Kolmogorov--Zakharov (KZ) fixed points and their inertial spectra \cite{Zakharov1992,Nazarenko2011}. Discrete
WT complements this by classifying integrable interaction graphs \cite{Kartashova2007PRL,Kartashova2009EPL}.
In kinetic WT, interaction topology enters through the first non-vanishing collision integral \cite{Zakharov1967Capillary}: $I_3$ for triads (CW) and $I_4$ for quartets (GW), while discrete WT identifies the
corresponding resonant clusters \cite{Kartashova2007PRL}.

The present Wilson--Fisher (WF) framework operates at a different level (Fig. \ref{fig:EndMatter_Fig4}): it does not replace kinetic or cluster theories, but organizes their inviscid fixed points into RG attraction basins using experimentally accessible coordinates $[\bar{\Omega}_\nu(N,y_g),\bar{\Sigma}_\nu(N,B_N)]$. These are orthogonal RG observables: $\bar\Omega_\nu$ measures the extent of the
inertial basin (UV reach), while $\bar\Sigma_\omega$ measures its depth (response strength). Both are governed by the same
interaction topology $N$, but probe distinct pieces of the RG physics: $\bar\Omega_\nu(N,y_g)$ is controlled by relevance and the viscous exit (linear RG/cutoff physics; Fig.~2 probes where the RG flow stops), whereas $\bar\Sigma_\omega(N,B_N)$ is controlled
by nonlinear saturation and response (Wegner scaling; Fig.~3 probes how the RG flow proceeds).

Looking forward, DSWT offers a controlled route from discrete to statistical turbulence: increasing resonance
broadening relative to mode spacing (or broadening the forcing band / enlarging the system) makes resonant clusters
overlap and the effective interaction topology becomes statistical. In that limit, the present topology-controlled
WF flows are expected to cross over toward the functional renormalization associated with fully developed Navier–Stokes turbulence \cite{EyinkGoldenfeld1994,Canet2022FRGTurbulence}, positioning DSWT as an experimental bridge between discrete and continuous nonequilibrium criticality.

\end{document}


\title{Supplementary Material for
``Wilson--Fisher Renormalization of Discrete Surface-Wave Turbulence''}

\author{Jos\'e A. Santiago}
\author{Mikheil Kharbedia}
\author{Basilio J. Garc\'ia}
\author{Francisco Monroy}

\date{\today}
\maketitle

This Supplementary Material provides the scaling arguments and renormalization-group (RG) considerations underlying the main text. In particular, we justify the definition of the running coupling, the Wilson--Fisher beta function, the topology-dependent nonlinear transfer rate, and we provide additional steps supporting the Reynolds-number scaling of the integrated inertial spectral weight reported in the main text.

\section{Running coupling and inertial spectra}

We analyze the surface-velocity signal $v(t)$ measured by laser Doppler vibrometry through its power spectral density $S(\omega)$,
defined by $\langle v^2\rangle=\int_0^\infty S(\omega)\,d\omega$.
In the inertial interval, weak wave turbulence predicts Kolmogorov--Zakharov (KZ) power laws
$S(\omega)\sim \omega^{-p}$, with $p=17/6$ for capillary waves and $p=5/2$ for gravity waves
\cite{Zakharov1992,Nazarenko2011}.

At scaling level, the velocity variance carried by a logarithmic band around $\omega$ is
$v_\omega^2\sim S(\omega)\,\Delta\omega$ with $\Delta\omega\sim\omega$, hence
\begin{equation}
v_\omega^2 \sim \omega\,S(\omega).
\label{eq:SM_amp}
\end{equation}
Using $v_\omega\sim \omega a_\omega$ to relate velocity to surface-displacement amplitude $a_\omega$ then gives
\begin{equation}
a_\omega^2 \sim \frac{S(\omega)}{\omega}.
\label{eq:SM_a_from_v}
\end{equation}

\section{Dimensionless nonlinearity and RG coupling}

For weakly nonlinear surface waves, the standard control of nonlinearity is the wave steepness
$\epsilon(\omega)\sim k(\omega)a_\omega$, which measures the relative slope of the interface and sets the size of nonlinear
frequency shifts and resonance broadening; accordingly, it is the small parameter underlying weak-turbulence closures
\cite{Zakharov1992,Nazarenko2011,DiasKharif1999}. Combining with Eq.~(\ref{eq:SM_a_from_v}) yields
\begin{equation}
\epsilon(\omega)^2 \sim k(\omega)^2\,\frac{S(\omega)}{\omega}.
\label{eq:SM_steep}
\end{equation}
This motivates the RG running coupling used throughout the main text,
\begin{equation}
g(\omega)\equiv \epsilon(\omega)^2,
\label{eq:SM_g_def}
\end{equation}
which is dimensionless, scale dependent, and small in the DSWT regime.

\section{Wilson--Fisher beta function}

Introducing the RG ``time'' $\ell=\ln(\omega/\omega_0)$, the inertial KZ--scaling and the dispersion relation $k(\omega)\sim\omega^\alpha$ imply
\begin{equation}
g(\omega)\sim \omega^{\,y_g},
\qquad
y_g = 2\alpha-(p+1),
\label{eq:SM_yg}
\end{equation}
where we used $g(\omega)=\epsilon(\omega)^2\sim k(\omega)^2 S(\omega)/\omega$ [Eq.~(\ref{eq:SM_steep})] and $S(\omega)\sim\omega^{-p}$.
At scaling level, the RG flow of the coupling may be written in Wilson--Fisher form,
\begin{equation}
\beta(g)\equiv\frac{dg}{d\ell}
=
y_g\,g - B_N\,g^{\,N-1},
\label{eq:SM_beta}
\end{equation}
where $N$ is the order of the resonant interaction ($N=3$ for capillary-wave triads and $N=4$ for gravity-wave tetrads), and $B_N>0$ encodes nonlinear saturation. Besides the unstable Gaussian fixed point $g^\star=0$, Eq.~(\ref{eq:SM_beta}) admits a nontrivial stable fixed point corresponding to the KZ inertial state discussed in the main text.

\section{Topology-dependent nonlinear transfer rate}

In wave turbulence theory, spectral transfer is mediated by resonant $N$-wave interactions. The interaction vertex is of order $(N-2)$ in the wave amplitude, and the collision integral governing spectral evolution is quadratic in this vertex. As a result, the nonlinear transfer rate scales as
\begin{equation}
\tau_{\mathrm{nl}}^{-1}(\omega)\sim \omega\,\epsilon(\omega)^{2(N-2)} .
\label{eq:SM_taunl_eps}
\end{equation}
Substituting $\epsilon(\omega)^2\sim k(\omega)^2 S(\omega)/\omega$ [Eq.~(\ref{eq:SM_steep})] into Eq.~(\ref{eq:SM_taunl_eps}) yields
\begin{equation}
\tau_{\mathrm{nl}}^{-1}(\omega)
\sim
\omega\left[\frac{k(\omega)^2}{\omega}\,S(\omega)\right]^{\,N-2},
\label{eq:SM_taunl_spec}
\end{equation}
which reproduces the bridge relation used in the main text in velocity-PSD form and makes explicit that the dependence on interaction topology enters solely through the exponent $N-2$.

\section{Matching to viscous dissipation: Kolmogorov cutoff scaling (Support for Figure~2)}

Viscous damping originates from the Navier--Stokes operator $\nu\nabla^2\bm{u}$ and yields, at scaling level, a universal dissipation rate
\begin{equation}
\tau_\nu^{-1}(k)\sim \nu k^2 .
\label{eq:SM_visc_rate}
\end{equation}
The ultraviolet cutoff frequency is experimentally identified as the Kolmogorov cutoff $\omega_K$ (Fig.~1 of the main text) and its renormalized scaling with forcing is reported in Fig.~2. It is selected by the crossover condition
\begin{equation}
\tau_{\mathrm{nl}}(\omega_K)\sim\tau_\nu(\omega_K),
\label{eq:SM_matching}
\end{equation}
which terminates the RG flow at finite scale.

To compare different fluids and dispersive regimes, we normalize the exit frequency by the viscous decay rate of the \emph{injection} mode. Let $k_0\equiv k(\omega_0)$ be the driven wavenumber selected by the dispersion relation at the fixed forcing frequency $\omega_0$. We define the dimensionless ultraviolet coordinate
\begin{equation}
\bar \Omega_\nu \equiv \frac{\omega_K}{\nu k_0^2}\;=\;\frac{\text{exit rate}}{\text{viscous decay rate at injection}},
\label{eq:SM_Omeganu_def}
\end{equation}
which measures how far the cascade extends before being arrested by viscous dissipation, in units of the injection-scale viscous rate $\tau_\nu^{-1}(k_0)\sim\nu k_0^2$. This normalization does not require $k_0$ to be invariant across fluids; it relies only on the microscopic Navier--Stokes operator evaluated at the forced mode.

The definition~(\ref{eq:SM_Omeganu_def}) cleanly separates the RG-generated ultraviolet extent of the cascade, set by the balance~(\ref{eq:SM_matching}) at the exit scale, from the bare dissipative rate fixed at injection. Changing the precise numerical definition of $k_0$ by an $\mathcal{O}(1)$ factor rescales $\bar\Omega_\nu$ by a constant prefactor but leaves the measured power-law exponents unchanged. The observed Wegner-like crossover scaling in Fig.~2 therefore reflects a genuine Wilsonian property of the viscous cutoff, not an artifact of a particular injection-scale convention. Using the topology-dependent transfer rate in velocity-PSD form,
\begin{equation}
\tau_{\mathrm{nl}}^{-1}(\omega)\sim \omega\!\left[\frac{k(\omega)^2}{\omega}\,S(\omega)\right]^{N-2},
\label{eq:SM_taunl_velPSD}
\end{equation}
(see Eq.~(\ref{eq:SM_taunl_spec})), the cutoff inherits its forcing dependence through the inertial KZ spectrum
$S(\omega)\sim \Pi^{\,\chi}\omega^{-p}$, where $\Pi$ is the conserved energy flux and $\chi$ is fixed by weak-turbulence theory
\cite{Zakharov1992,Nazarenko2011}. At fixed driving frequency $\omega_0$, the injected kinetic-energy density
$E=\tfrac12\rho U_0^2$ sets the flux at scaling level, $\Pi\propto E$ (up to $\omega_0$-dependent constants).
Combining these scalings with $k(\omega)\sim\omega^\alpha$ in the matching condition~(\ref{eq:SM_matching}) yields the
topology-dependent cutoff laws reported in the main text and validated by the collapse in Fig.~2,
\begin{equation}
\bar\Omega_{\nu}^{(C)} \sim \bar E^{2/3}\quad (N=3),
\qquad
\bar\Omega_{\nu}^{(G)} \sim \bar E^{1}\quad (N=4).
\label{eq:SM_cutoff}
\end{equation}

For each forcing series performed at fixed $\omega_0$, the ultraviolet reach may equivalently be parametrized by
$\bar\Omega_{\omega}\equiv \omega_K/\omega_0$; however, $\bar\Omega_\nu$ is the more universal normalization across fluids,
as it explicitly removes the fluid-dependent viscous rate at injection. In this sense, $\bar\Omega_\nu$ is a genuinely
Wilsonian ultraviolet coordinate: it compares the RG-generated exit scale $\omega_K$ to the bare dissipative operator inherited from microscopic hydrodynamics. The reduced forcing is expressed as $\bar E\equiv E/E_0$, with $U_0=A\omega_0$ and viscous reference energy
\begin{equation}
E_0 \equiv \rho\left(\frac{\nu}{\Lambda_0}\right)^2,
\qquad
\Lambda_0\equiv\frac{2\pi}{k_0}.
\label{eq:SM_E0_def}
\end{equation}
Here $\Lambda_0$ is the injection wavelength selected by the linear dispersion relation at $\omega_0$.
Physically, $E_0$ is the kinetic-energy density associated with viscous relaxation over one forcing wavelength, so that
$\bar E\sim (U_0\Lambda_0/\nu)^2$ measures the distance from the viscous (Gaussian) regime using only injection-scale
quantities, isolating the topology-controlled renormalization of the ultraviolet cutoff.

\section{Integrated inertial spectral weight: Reynolds scaling (support for Figure~3)}

We connect the experimentally accessible integrated inertial spectral weight
\begin{equation}
\Sigma_{\mathrm{PSD}} \equiv \int_{\omega_0}^{\omega_K} S(\omega)\,d\omega,
\label{eq:SM_Sigma_def}
\end{equation}
to the Reynolds-number scaling reported in Fig.~3 of the main text. Here $S(\omega)$ is the surface-velocity PSD measured by LDV, so $\Sigma_{\mathrm{PSD}}$ has units of velocity squared and represents the total velocity variance accumulated in the direct cascade between the pump and the ultraviolet exit scale. The upper limit $\omega_K$ is retained because it marks the experimentally resolved exit from the inertial basin; however, for the steep inertial spectra observed here ($p>1$), extending the integral to $\infty$ does not affect scaling exponents (the integral is pump-dominated).

\subsection*{1. Pump-dominated integral and flux prefactor}

For a stationary direct energy-flux cascade, the inertial KZ spectrum takes the scaling form
\begin{equation}
S(\omega)\sim C_N\,\Pi^{\,\xi}\,\omega^{-p},
\label{eq:SM_KZ_flux}
\end{equation}
where $\Pi$ is the constant energy flux through the inertial interval and $\xi$ is fixed by weak-turbulence scaling for the relevant $N$-wave kinetic equation \cite{Zakharov1992,Nazarenko2011}. Substituting Eq.~(\ref{eq:SM_KZ_flux}) into Eq.~(\ref{eq:SM_Sigma_def}) gives, for $p\neq 1$,
\begin{equation}
\Sigma_{\mathrm{PSD}}
\sim
\Pi^{\,\xi}\,\frac{\omega_K^{1-p}-\omega_0^{1-p}}{1-p}.
\label{eq:SM_Sigma_int}
\end{equation}
Because $p>1$ in both classes, the integral is dominated by its lower limit, so at scaling level
\begin{equation}
\Sigma_{\mathrm{PSD}}\sim \Pi^{\,\xi}\,\omega_0^{\,1-p}\qquad (p>1),
\label{eq:SM_Sigma_Pi}
\end{equation}
with $\omega_K$ contributing only a subleading correction $\propto \Pi^{\,\xi}\omega_K^{1-p}$.

\subsection*{2. Reduced response coordinate and Reynolds control}

We define as reduced (dimensionless) cascade response the forcing-normalized integrated variance
\begin{equation}
\bar\Sigma_{\omega} \equiv \frac{\Sigma_{\mathrm{PSD}}}{(\Lambda_0\omega_0)^2}
=
\frac{\text{inertial velocity variance above the pump}}
     {\text{forcing-scale velocity squared}},
\qquad 
\Lambda_0\equiv\frac{2\pi}{k_0},
\label{eq:SM_Sigma_bar}
\end{equation}
where $k_0$ is the driven wavenumber selected by the dispersion relation at the fixed forcing frequency $\omega_0$.
This is the appropriate response coordinate for Fig.~3 because it removes only the trivial forcing-scale units and leaves viscosity
to enter solely through renormalization. A viscous normalization,
$\Sigma_{\mathrm{PSD}}/(\nu/\Lambda_0)^2$, would divide by the relevant perturbation itself and hard-wire an explicit $\nu^{-2}$ factor,
masking the crossover physics.

In the pump-dominated regime $p>1$, Eqs.~(\ref{eq:SM_Sigma_Pi})--(\ref{eq:SM_Sigma_bar}) imply, at fixed $\omega_0$ and $\Lambda_0$,
\begin{equation}
\bar\Sigma_{\omega} \sim \Pi^{\,\xi},
\label{eq:SM_Sigma_flux_control}
\end{equation}
so the Reynolds dependence of $\bar\Sigma_{\omega}$ reflects how viscosity renormalizes the inertial flux (equivalently the injection-scale coupling).

The experimentally accessible control parameter is the injection Reynolds number
\begin{equation}
\mathrm{Re}=\frac{U_0\Lambda_0}{\nu},
\qquad U_0=A\omega_0.
\label{eq:SM_Re_def}
\end{equation}

\subsection*{3. WF/Wegner scaling and quadratic response to the WF field}

The Wilson--Fisher (WF) framework promotes the squared steepness $g(\omega)=\epsilon(\omega)^2$ to a running coupling whose flow is controlled by
\begin{equation}
\beta(g)=\frac{dg}{d\ell}=y_g g - B_N g^{\,N-1},
\qquad \ell\equiv \ln(\omega/\omega_0),
\label{eq:SM_beta_recall}
\end{equation}
with interaction topology fixing the first nonlinear saturation channel $g^{\,N-1}$ (triads: $N=3$ for CW; tetrads: $N=4$ for GW).
The reduced response follows a Wegner crossover form,
\begin{equation}
\bar\Sigma_{\omega}(\mathrm{Re}) \;=\; \mathrm{Re}^{\,\kappa_N}\,
\mathcal{F}_N\!\left(\frac{\mathrm{Re}}{\mathrm{Re}_\times}\right),
\label{eq:SM_Wegner_Sigma}
\end{equation}
where $\mathcal{F}_N$ is topology dependent and ${\rm Re}_\times$ is nonuniversal. The key closure supported by Fig.~3 is that the
forcing-normalized integrated variance is quadratic in the coarse-grained WF field at injection,
\begin{equation}
\bar\Sigma_\omega \;\propto\; g_0^{\,2},\qquad g_0\equiv g(\omega_0).
\label{eq:SM_Sigma_g0_quadratic}
\end{equation}
This dependence is dynamical: in weak WT the transfer amplitude is proportional to the interaction vertex and hence to the steepness
$\epsilon$, so the leading variance production above the pump scales as the square of that amplitude, $\sim\epsilon^2$. Since
$g=\epsilon^2$, the leading cascade-induced contribution scales as $\delta\bar\Sigma_\omega\sim g_0^2$, consistent with the lowest
non-vanishing self-renormalizations summarized in End Matter, Fig.~4.

\subsection*{4. Topology-controlled Reynolds scaling of the cascade response}

In the asymptotic WF regime (${\rm Re}\gg{\rm Re}_\times$), the crossover function in
Eq.~(\ref{eq:SM_Wegner_Sigma}) saturates, so $\bar\Sigma_\omega\sim{\rm Re}^{\kappa_N}$.
To determine $\kappa_N$, we write the injection-scale effective coupling as the single-parameter
crossover ansatz $g_0\sim{\rm Re}^{\gamma_N}$. The exponent $\gamma_N$ is fixed by topology:
the first nonlinear feedback that arrests the WF growth is the $g^{\,N-1}$ term in
Eq.~(\ref{eq:SM_beta_recall}), so the minimal injection-level equation of state is
${\rm Re}\propto g_0^{\,N-1}$, i.e.
\begin{equation}
g_0 \sim {\rm Re}^{\,1/(N-1)} .
\label{eq:SM_g0_Re}
\end{equation}
Combining Eq.~(\ref{eq:SM_g0_Re}) with the quadratic response law
Eq.~(\ref{eq:SM_Sigma_g0_quadratic}) gives
\begin{equation}
\bar\Sigma_\omega \sim {\rm Re}^{\,2/(N-1)} ,
\label{eq:SM_Sigma_Re_general}
\end{equation}
hence $\kappa_N=2/(N-1)$. Therefore,
\begin{equation}
\bar\Sigma_\omega^{(CW)}\sim {\rm Re}^{1}\quad (N=3),
\qquad
\bar\Sigma_\omega^{(GW)}\sim {\rm Re}^{2/3}\quad (N=4),
\label{eq:SM_Sigma_final_Re}
\end{equation}
in agreement with the collapse in Fig.~3 of the main text.

\section{Raw integrated inertial spectral weight (Supplementary Fig.~S1)}

We report the non-normalized integrated inertial spectral weight
\begin{equation}
\Sigma_{\mathrm{PSD}} \equiv \int_{\omega_0}^{\omega_K} S(\omega)\, d\omega ,
\label{eq:SM_Sigma_raw_def}
\end{equation}
where $S(\omega)$ is the LDV surface-velocity PSD and $\omega_K$ is the experimentally resolved viscous cutoff. Since
$\Sigma_{\mathrm{PSD}}$ has units of velocity squared, it measures the total velocity variance accumulated in the direct
cascade between the pump and the ultraviolet exit scale.

Supplementary Fig.~S1 plots $\Sigma_{\mathrm{PSD}}$ versus the forcing velocity $U_0=A\omega_0$ for gravity-wave (GW) and
capillary-wave (CW) cascades across all fluids. In both classes, $\Sigma_{\mathrm{PSD}}$ increases monotonically with forcing,
showing direct activation of the inertial response above the pump. The raw curves are not expected to collapse because
$\Sigma_{\mathrm{PSD}}$ retains explicit dependence on the forcing scale and material parameters; the corresponding reduced
response $\bar\Sigma_\omega$ and its Reynolds renormalization are reported in Fig.~3 of the main text.

\begin{figure}[htb!]
    \centering
    \includegraphics[width=\columnwidth]{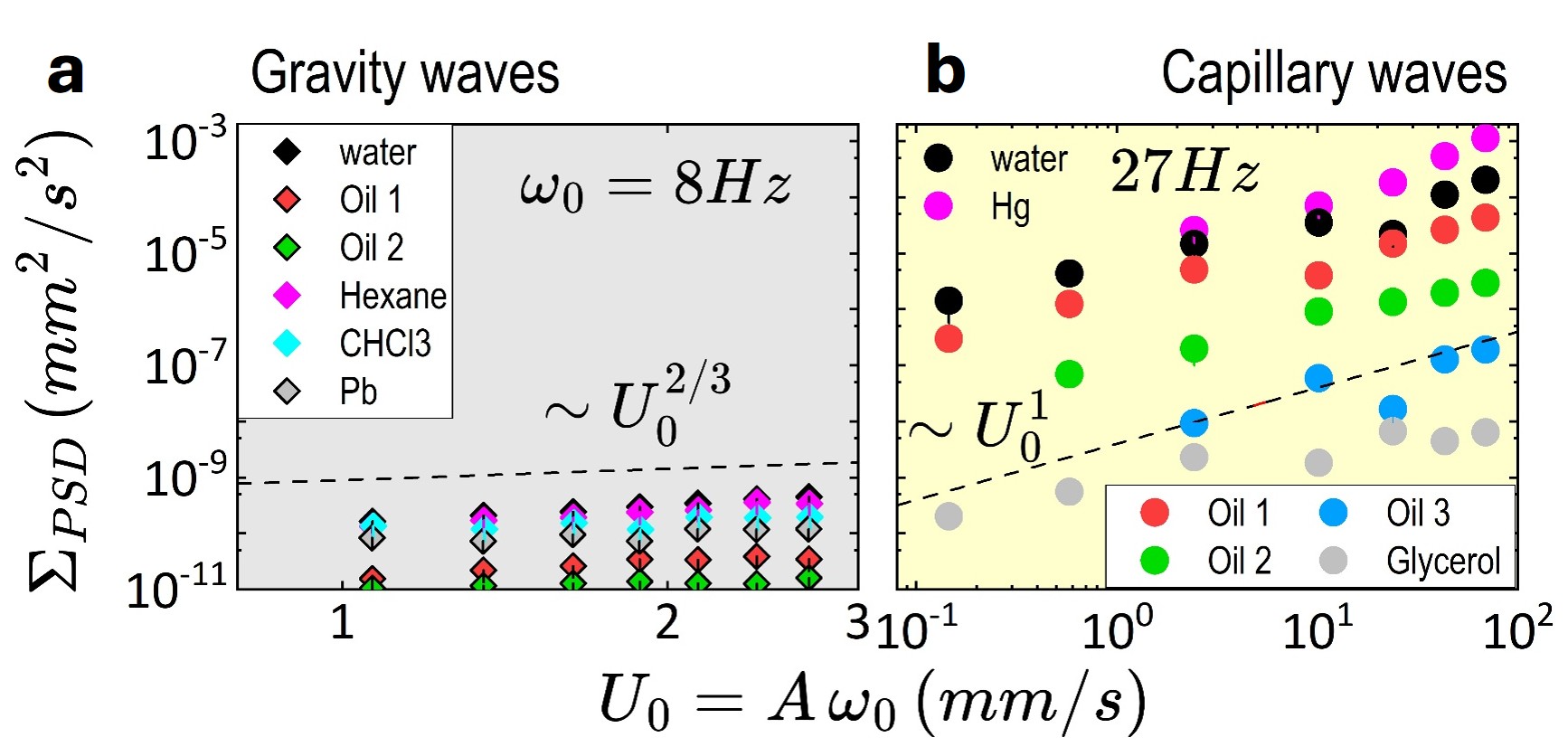}
    \caption{
    \textbf{Supplementary Fig.~S1: Raw integrated inertial spectral weight.}
    Integrated inertial spectral weight $\Sigma_{\mathrm{PSD}}=\int_{\omega_0}^{\omega_K} S(\omega)\,d\omega$
    (LDV surface-velocity PSD) versus forcing velocity $U_0=A\omega_0$ for gravity waves (left; $\omega_0=8\,\mathrm{Hz}$)
    and capillary waves (right; $\omega_0=27\,\mathrm{Hz}$), across all fluids (legend).
    The GW branch lies $\sim 3$ decades below the CW branch over the full forcing range.
    The forcing-normalized response $\bar\Sigma_\omega\equiv \Sigma_{\mathrm{PSD}}/(\Lambda_0\omega_0)^2$ and its Reynolds
    renormalization are shown in Fig.~3 of the main text.
    }
    \label{fig:S1_rawSigma}
\end{figure}

\section*{Orthogonality of RG coordinates: exit scale versus response}

The two reduced observables used in the main text, the exit coordinate $\bar\Omega_\nu$ (Fig.~2) and the cascade
response $\bar\Sigma_\omega$ (Fig.~3), probe complementary pieces of the WF structure. The exit coordinate $\bar\Omega_\nu=\omega_K/(\nu k_0^2)$ is normalized by the viscous decay rate of the injection mode,
so it measures the ultraviolet reach of the inertial basin relative to the bare Navier--Stokes dissipative operator.
By contrast, $\bar\Sigma_\omega=\Sigma_{\mathrm{PSD}}/(\Lambda_0\omega_0)^2$ is normalized only by forcing-scale quantities and
measures the strength of the cascade, i.e.\ the velocity variance accumulated above the pump; using a viscous normalization
here would mix the response with the relevant perturbation that drives the crossover. This asymmetric normalization is deliberate: $\bar\Omega_\nu$ quantifies where the flow stops (viscous arrest), whereas
$\bar\Sigma_\omega$ quantifies how strongly it proceeds (nonlinear saturation). Together, $(\bar\Omega_\nu,\bar\Sigma_\omega)$ form
an orthogonal pair of RG coordinates that separately resolve reach and intensity in DSWT.

\section{Scope of validity}

The arguments presented here are valid at the scaling level within the weakly nonlinear DSWT regime, where
(i) the inertial interval exhibits KZ-type power-law behavior,
(ii) transfer is mediated predominantly by the resonant interaction topology ($N=3$ for CW and effective $N=4$ for GW),
and (iii) viscous damping can be represented at scaling level by $\tau_\nu^{-1}(k)\sim \nu k^2$.
The crossover matching $\tau_{\mathrm{nl}}(\omega_K)\sim\tau_\nu(\omega_K)$ is intended as a scaling criterion for the exit from the inertial basin.

\subsection*{Weak-nonlinearity diagnostic}

We monitored the forcing-scale steepness $\epsilon_0 \equiv k_0 A$ as a dimensionless proxy for nonlinear frequency shifts and resonance broadening.
Across all datasets reported here, $\epsilon_0$ remains in the range
\[
10^{-3} \;\lesssim\; \epsilon_0 \;\lesssim\; 10^{-1},
\]
well below unity and firmly within the weakly nonlinear regime. In this range, the pump remains spectrally sharp, higher harmonics are subdominant, and a clear KZ-type inertial interval is resolved above $\omega_0$.

As $\epsilon_0$ approaches unity, nonlinear frequency shifts become comparable to linear mode spacing, leading to strong resonance broadening, cluster overlap, and eventual breakdown of topology-controlled transfer. Such conditions are expected to invalidate the present Wilson--Fisher crossover description and drive a transition toward strong (Philips-like) turbulence, a regime not explored in the present experiments.